\documentclass[11pt]{scrartcl}
\usepackage[margin=3cm
]{geometry}

\usepackage{graphicx}% Include figure files
\usepackage{dcolumn}% Align table columns on decimal point
\usepackage{bm}% bold math
\usepackage{placeins}
\usepackage{hyperref}% add hypertext capabilities
\usepackage{authblk}
\newcommand{\fig}[5]{
	\begin{figure}
	\centerline{\includegraphics[width=#5\textwidth]{#1}}
\caption[#3]{\label{#2} \textbf{#3} #4}
\end{figure}}

\begin{document}

\title{Electric coupling in scanning SQUID measurements}

%\author{Eric M. Spanton\textsuperscript{1,2}, Aaron J. Rosenberg\textsuperscript{3}, , Kathryn A. Moler\textsuperscript{1,2,3}}
% \affiliation{Stanford University}

\author[1,2]{Eric M. Spanton}
\author[3]{Aaron J. Rosenberg}
\author[1,2,3]{Yihua H. Wang}
\author[3]{John R. Kirtley}
\author[4,5]{Ferhat Katmis}
\author[5]{Pablo Jarillo-Herrero}
\author[4]{Jagadeesh S. Moodera}
\author[1,2,3,*]{Kathryn A. Moler}

\affil[1]{Stanford Institute for Materials and Energy Sciences, SLAC National Accelerator Laboratory, Menlo Park, CA, USA}
\affil[2]{Department of Physics, Stanford University, Stanford, CA, USA}
\affil[3]{Department of Applied Physics, Stanford University, Stanford, CA, USA}
\affil[4]{Francis Bitter Magnet Laboratory and Plasma Science and Fusion Center, Massachusetts Institute of Technology, Cambridge, MA, USA}
\affil[5]{Department of Physics, Massachusetts Institute of Technology, Cambridge, MA , USA}
\affil[*]{Corresponding Author: kmoler@stanford.edu}
\renewcommand\Authands{ and }

\date{\today}% 

\maketitle
\begin{center}
\textit{Note: A retraction of Ref. \cite{Wang2015} will appear in Science on December 18th, 2015. Here we explain in detail the reasons for the retraction.}
\end{center}

\begin{abstract}
Scanning SQUID is a local magnetometer which can image flux through its pickup loop due to DC magnetic fields ($\Phi$). Scanning SQUID can also measure a sample's magnetic response to an applied current ($d\Phi/dI$) or voltage ($d\Phi/dV$) using standard lock-in techniques. In this manuscript,  we demonstrate that electric coupling between the scanning SQUID and a back gate-tuned, magnetic sample can lead to a gate-voltage dependent artifact when imaging $d\Phi/dI$ or $d\Phi/dV$. The electric coupling artifact results in $d\Phi/dV$ and $d\Phi/dI$  images which mimic the spatial variation of the static magnetic fields from the sample (e.g. ferromagnetic domains). In back-gated $EuS/Bi_2Se_3$ bilayers, we show that the electric coupling effect is important, and is responsible for the reported signal from chiral currents in Wang et al. \cite{Wang2015}. Previous scanning SQUID current imaging experiments are unaffected by this artifact, as they are either on non-magnetic samples or the spatial distribution of magnetism does not match the features observed in $d\Phi/dI$. In conclusion, $d\Phi/dI$ or $d\Phi/dV$ imaging of magnetic, back-gated samples should only be applied and interpreted with great caution.
\end{abstract}

% {\let\thefootnote\relax\footnotetext{\textsuperscript{1} \textit{Stanford Institute for Materials and Energy Sciences, SLAC National Accelerator Laboratory, Menlo Park, California 94025, USA}}}

%{\let\thefootnote\relax\footnotetext{\textsuperscript{2} \textit{Department of Physics, Stanford University, Stanford, California 94305, USA}}}

%{\let\thefootnote\relax\footnotetext{\textsuperscript{3} \textit{Department of Applied Physics, Stanford University, Stanford, California 94305, USA}}}

\newpage

\section{Executive Summary}

In Ref.\cite{Wang2015}, written by several of us, scanning superconducting quantum interference devices (SQUID) were used to investigate the magnetic domain structure and current flow in $EuS/Bi_2Se_3$ bilayers. A back gate voltage was used to tune the chemical potential of the $Bi_2Se_3$. At negative gate voltages, images of flux from current applied to the sample ($d\Phi/dI$) showed features which were interpreted as signatures of domain wall currents associated with the quantum anomalous Hall effect (QAHE). The signal in $d\Phi/dI$ (from written magnetic domain structure) scaled with the local voltage of the sample, rather than the current, which was interpreted as evidence of the current's chiral nature.

In the section \ref{model}, \textbf{Sample-SQUID electric coupling model}, we derive a model for electric coupling between the SQUID and sample, which leads to an artifact in $d\Phi/dI$ and $d\Phi/dV$ imaging when the sample is back-gated and magnetic. The charge of the back-gated $Bi_2Se_3$ device is modulated when a voltage is applied to the sample. The charge of the sample couples electrically to the grounded metal on the SQUID itself. The electric coupling causes the height of the SQUID's pickup loop above the sample to oscillate when an oscillating voltage or current is applied to the sample. In the presence of magnetic fields from the sample, the gradient of the local magnetic field is coupled into the $d\Phi/dI$ and $d\Phi/dV$ images by the oscillation of the SQUID. We simulated $d\Phi/dI$ images from both chiral currents and the electric coupling artifact, which is proportional to the height derivative of the magnetic image ($d\Phi/dz$). The images from chiral currents and the electric coupling artifact are qualitatively very similar, although the electric coupling artifact is noticeably sharper.

In section \ref{exp_signatures}, \textbf{Experimental signatures of electric coupling in $EuS/Bi_2Se_3$}, we compare the theoretical predictions of the electric coupling artifacts to new measurements of the field, frequency, and back gate voltage dependencies of magnetic images of $EuS/Bi_2Se_3$ bilayers. We found that current and voltage images ($d\Phi/dI$ and $d\Phi/dV$) match the expected features of the electric coupling artifact. Specifically, we imaged $d\Phi/dV$ and found that the sign of the signal reverses sign as a function of frequency and back gate voltage, and the signal disappears when the SQUID is in contact with the sample. These features are not consistent with a signal from chiral currents along domain walls.

Based on these measurements, we've reevaluated the $d\Phi/dI$ images in Ref. \cite{Wang2015} in \textbf{Implications for Ref. \cite{Wang2015}} (section \ref{implications}). We show that the detailed spatial dependence of $d\Phi/dI$ images taken by Wang et al. along the edge of a device (Fig. 2 in Ref. \cite{Wang2015}) very closely matches the height derivative of the magnetic image ($d\Phi/dz$), as expected for electric coupling. We also show previously unpublished data taken by Wang et al., showing that the $d\Phi/dI$ features observed in Fig. 2 disappear halfway through a scan, suggesting that the SQUID is in contact. Additionally, sign reversal of $d\Phi/dI$ features as a function of back gate are shown for a written domain.

The results of the new measurements and the presence of strong evidence for the electric coupling artifact in measurements performed for Ref. \cite{Wang2015} lead to the conclusion that the results of Ref. \cite{Wang2015} are primarily, if not completely due to electric coupling, rather than chiral currents.

Finally, in \textbf{Electric coupling in other current imaging experiments} (section \ref{other_works}) we briefly comment on other works where scanning SQUID was used to image current, and describe why the electric coupling artifact present in $EuS/Bi_2Se_3$ SQUID measurements are not responsible for the features observed in those works (\cite{kalisky,nowack,spanton}).

\section{Sample - SQUID electric coupling model} \label{model}
\FloatBarrier

\subsection{Model of electric coupling}
Scanning SQUID is a sensitive local flux to voltage converter. The SQUID's pickup loop is brought very close to a sample and scanned in order to image magnetic fields. Our SQUIDs also have a field coil which is concentric with the pickup loop, and can be used to apply a local field. Our SQUID devices are fabricated on silicon chips, which are polished close to the pickup loop in order to bring it very close to the sample (typically within a few microns). 

Our scanning SQUID chip is typically mounted on a copper cantilever (FIG.\ref{schematic} a) for scanning. We detect a deflection of this cantilever capacitively when the SQUID chip touches the sample, which gives us topographic imaging of the sample. We can either scan with the SQUID "in contact" with the tip of the polished SQUID chip touching the sample, or "out of contact" which means that the SQUID's tip scans above the sample. 

 \fig{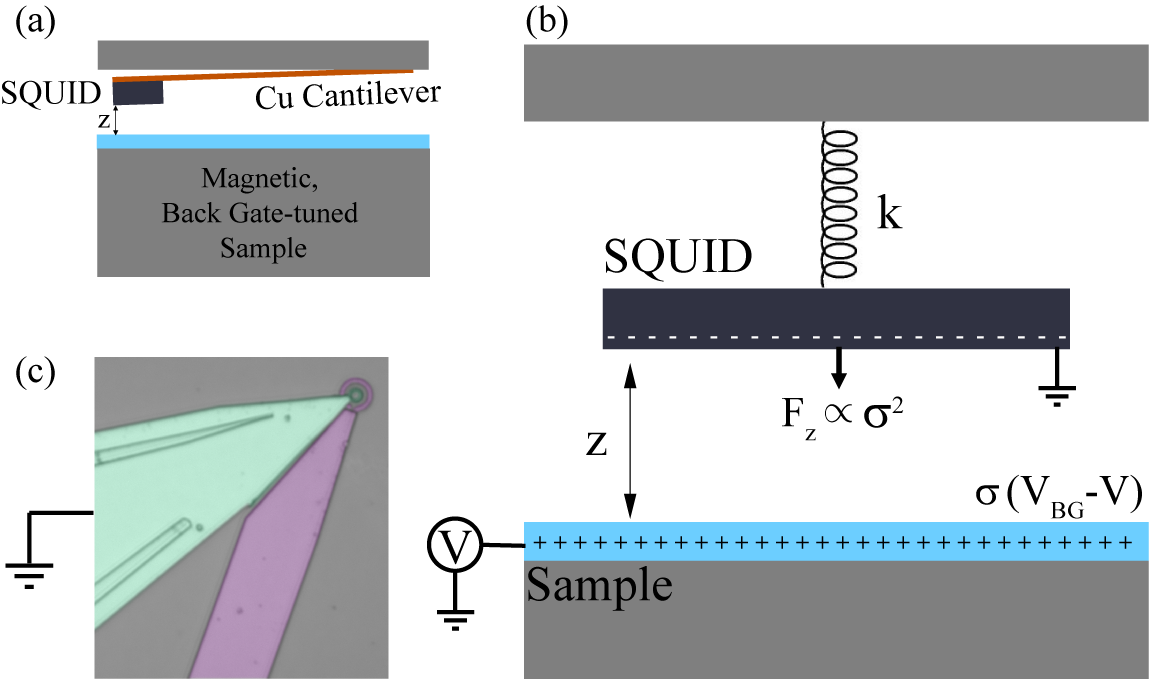}{schematic}{Overview of scanning SQUID measurements and a simple model of electric coupling.}
{(a) A schematic side view of a scanning SQUID measurement. The SQUID is attached to a copper cantilever, the deflection of which is used to determine when the SQUID is touching the sample. The height of the SQUID above the sample (z) is adjusted with piezo-based scanners (not shown). (b) A schematic of parameters relevant for the electric coupling model. The scanning SQUID's pickup loop, and often times the field coil as well, are electrically at ground during measurements. When measuring current (or in the case of a directly applied voltage) the sample is locally at some voltage, $V$. The SQUID is attached to a Cu cantilever which has a spring constant, k.The charge density on the sample is fixed by the sample's properties and the applied electric fields. In the case of a gated sample the net charge of the sample can be non-zero. The electric force between the grounded SQUID and the charge on the sample leads to a coupling between height and voltage of the sample (see text). (c) A false color image of a typical SQUID used for scanning, showing the SQUID's pickup loop and shielding (green) and the field coil and it associated shielding (purple).}{0.7}

When the SQUID is out of contact it is free to vibrate or deflect if a force is applied to it. We modeled the coupling between grounded metal on the SQUID and the charge on a gated sample (FIG. \ref{schematic} b). The tip of the scanning SQUID (FIG. \ref{schematic} c) has a pickup loop with superconducting shielding (green) and a field coil which also has shielding (purple). Both loops and their shielding are grounded during normal operation of the SQUID. A charge on the sample will exert a force on a grounded metallic SQUID. The Coulomb force will lead to a deflection of the SQUID which depends on the charge accumulated on the sample.

The following model is very similar to how charge is measured in electric force microscopy, however the amount of charge on the sample itself is also modulated by a voltage.

A generic electric equation for the force on the grounded SQUID due to a gated semiconducting sample with a voltage applied to the sample is:

\begin{equation}
F_z= a_1 \sigma^2 + a_2 (V-V_{CPD})^2 = a_1^* (V_{BG} - V)^2 + a_2 (V-V_{CPD})^2
\end{equation}

where $a_1$,$a_2$, and $a_1^*$ are parameters with the appropriate units which depend on the SQUID and sample geometry, $\sigma$ is the 2D charge density induced in the semiconductor by a back gate, $V$ is the voltage applied to the sample, $V_{BG}$ is the voltage applied to the back gate, and $V_{CPD}$ is the contact potential difference. 

The first term is the force between the charged sample and a grounded metallic object, and this is the term which is relevant for this paper. This force will be proportional to the square of the induced charge on the sample.  A second term, which is present due to the applied voltage difference between the sample and SQUID, leads to a term which goes as $V^2$. The contact potential difference (is a term which is important for Kelvin Probe Force microscopy \cite{kpfm}, for example) is typically less than 1 V in magnitude. 

The back gate acts as a parallel plate capacitor with the sample and back gate as the plates. Therefore, the induced charge density on the sample will be proportional to $V-V_{BG}$, leading to the second equality in Eqn. 1. For a lock-in measurement, we are only concerned with terms linear in V when we are measuring the 1st harmonic. If $V_{CPD}$ is small, the second term goes primarily as $V^2$ and will only show up in 2nd and higher harmonics. For the purposes of this discussion we will ignore this term, but it can in principle also lead to other artifacts.

We can now balance the electric force with the restoring force of the cantilever
 
 \begin{equation}
F_z = a_1^* (V_{BG} - V)^2 = k (z-z_0)
\end{equation}

Where k is the spring constant of the cantilever and $z_0$ is its equilibrium position. Solving for $z-z_0$, we find:

 \begin{equation}
\Delta z \equiv z-z_0 = \frac{a_1^*}{k}  (V_{BG}^2 + V^2 - 2 V_{BG} V)
\end{equation}

Again, since we only are concerned with terms linear in V for lock-in measurements, we can drop the first two terms and find the first harmonic of the response to a sinusoidal excitation:

 \begin{equation}
\Delta z  = -\frac{2 a_1^*}{k} V_{BG} V sin(\omega t)
\end{equation}

We have now established, for a back-gated sample with an induced charge and a grounded SQUID, that the application of a sinusoidal voltage excitation to the sample can lead to a sinusoidal height variation of the SQUID. This height variation directly couples the gradient of any DC magnetic fields in the sample into a lockin measurement of the SQUID's response.

By applying a voltage to the sample and measuring the SQUID's response, we are measuring the constant component of $\frac{d\Phi}{dV}$. The full measured $\frac{d\Phi}{dV}$, taking into account the above height variation is:

\begin{equation}
\frac{d\Phi}{dV} = \frac{\partial \Phi}{\partial V} + \frac{\partial \Phi}{\partial z} \frac{dz}{dV} = \frac{\partial \Phi}{\partial V}-\frac{2 a_1^*}{k} V_{BG} \frac{\partial \Phi}{\partial z}
\end{equation}

The first term is the true change in the flux from the sample due to the applied voltage, for example from a chiral current which is modulated by gate voltage or from voltage-induced domain wall motion. The second term is the 'electric coupling artifact' term due to linear coupling of the height of the SQUID to the applied voltage, which is the main result of this section.

Similarly, if we want to measure $\frac{d\Phi}{dI}$, there is an electric coupling term. Flowing a current in a resistive sample leads to a voltage drop across the sample. The local voltage of the sample  near the SQUID will induce a coupling to the height of the SQUID. The strength of the electric coupling ($\frac{dz}{dI}$) however, will vary as a function of position along the sample and the sample's resistance. In our model, the electric coupling at a fixed position will be proportional to R, ($\frac{dz}{dI} \propto R V_{BG}$).

The electrostatic model established here is simplistic. We have not included a realistic model of the actual form of the electric coupling, $a_1^*$ or the frequency dependence of the response of the cantilever. The electric coupling $a_1^*$, depends on the geometry and distance between the SQUID and sample and the screening from the sample itself. As the sample becomes more conducting, the screening of the back gate induced charge will decrease the electric coupling coefficient, which in the case of $EuS/Bi_2Se_3$ means that the electric coupling artifact will be diminished at positive gate voltages, where the sample is more conducting. Additionally, the resonance of the cantilever will also lead to a mechanical resonance-like response in the strength and sign of the electric coupling artifact.

In conclusion, the robust predictions of the electric coupling model presented here are that the electric coupling signal will be proportional to $d\Phi/dz$, it will change sign as a function of back gate voltage, and it will also be proportional to the mechanical response (and any resonances) of the cantilever.

\subsection{Signals due to chiral currents and electric coupling are qualitatively similar}

\fig{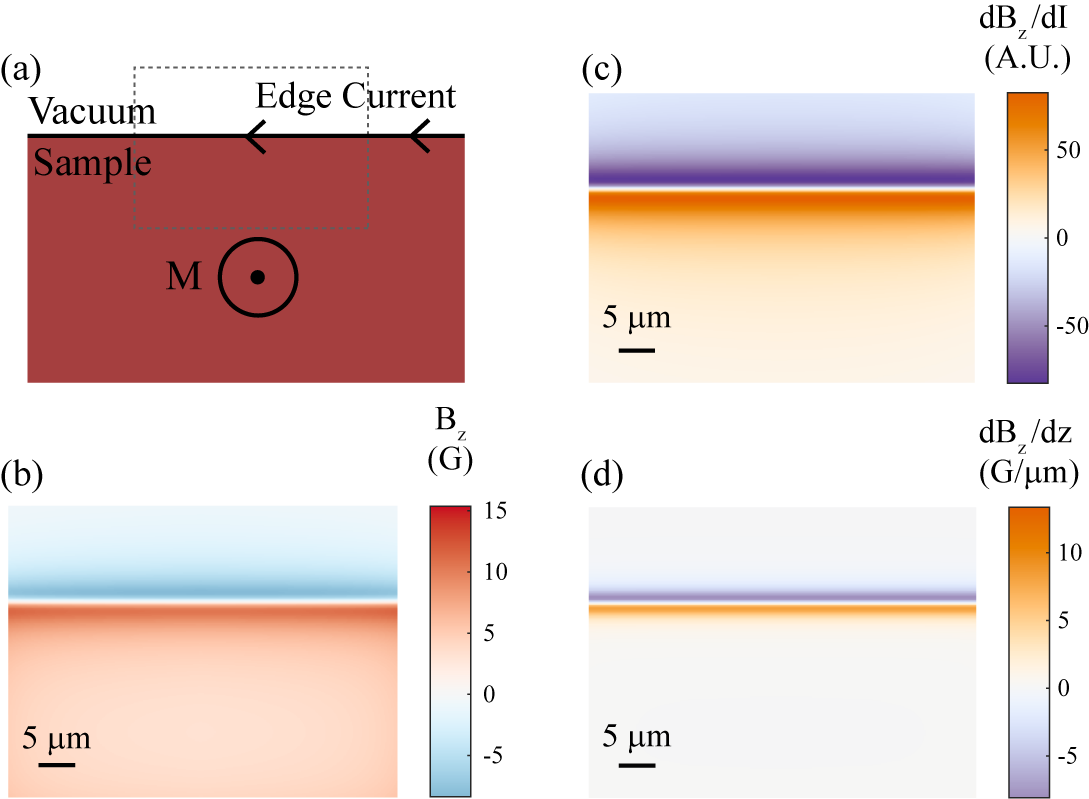}{edge_sim}{Simulated $B_z$, $\frac{dB_z}{dI}$, $\frac{dB_z}{dz}$ images at the edge of a polarized out-of-plane ferromagnet}
{(a) Simulated sample. (b) Simulated $B_z$ image at h=$1.5 \mu m$. (c) Simulated out of plane field ($B_z$) from a chiral current along the edge of the sample. (d) Simulated signal from the electric coupling artifact ($\frac{dB_z}{dz}$)}{0.7}

\fig{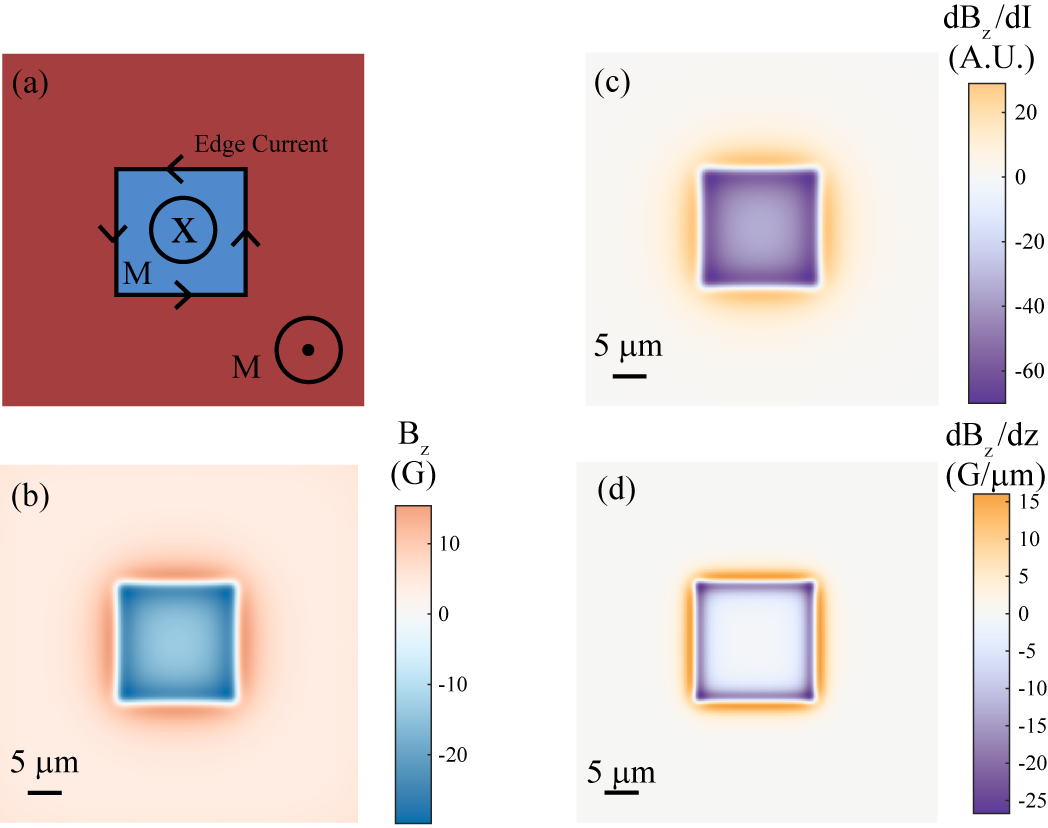}{domain_sim}{Simulated $B_z$, $\frac{dB_z}{dI}$, $\frac{dB_z}{dz}$ images of a written square magnetic domain}
{(a) Simulated square up domain written in a film which is polarized down everywhere else. (b) Simulated $B_z$ image at h=$1.5 \mu m$. (c) Simulated out of plane field ($B_z$) from a chiral currents along the domain wall. (d) Simulated signal from the electric coupling artifact ($\frac{dB_z}{dz}$)}{0.7}

In the quantum anomalous Hall state, chiral currents flow along the edges of devices and at domain walls \cite{qahe_review}. The magnetic fields (specifically the out of plane component, $B_z$) from chiral currents should be spatially resolvable by scanning SQUID microscopy. We simulated the expected $d\Phi/dI$ images due to currents along the edge of a device and at the wall of a written magnetic domain, and then compared them to the simulated image due to the electric coupling artifact derived above. We find that they are qualitatively similar, and therefore great caution must be exercised in attempting to measure chiral currents in samples with magnetic structure using scanning SQUID. 

We simulated images of the expected signals at the edge of a sample with a uniform out-of-plane polarization (FIG. \ref{edge_sim}) and around a written domain (FIG. \ref{edge_sim}). The expected signal from current flowing along the edge of the sample (\ref{edge_sim} c) is qualitatively reproduced by the electric coupling artifact, $dB_z/dz$  (FIG. \ref{edge_sim} d).
 Similarly, the current flowing along the domain wall of a written domain (FIG. \ref{domain_sim} c) looks qualitatively similar to the electric coupling artifact, $dB_z/dz$ (FIG. \ref{domain_sim} d).

The spatial dependence of the chiral current images and the electric coupling images are different in the details. The chiral current and magnetic field images for out of plane domains, however, are identical up to an overall scaling factor for the geometries we've simulated (FIG. \ref{domain_sim} b and c). Therefore, comparing images of $d\Phi/dI$ to both $\Phi$ and  $d\Phi/dz$ gives us another way of discriminating between real chiral current signals and electric coupling artifacts.

In the following section, we will use the SQUID-sample coupling model to show that electric coupling is dominant in new measurements we've performed on $EuS/Bi_2Se_3$.

\FloatBarrier

\section{Experimental signatures of electric coupling in $EuS/Bi_2Se_3$}\label{exp_signatures}
\FloatBarrier
Here we will give four pieces of evidence that strongly indicate that the electric coupling term is dominant in our new measurement of $EuS/Bi_2Se_3$ bilayers. We argue that all of the signals observed in this measurement of $d\Phi/dV$ and $d\Phi/dI$ are due to electric coupling. The four pieces of evidence are:

\begin{enumerate}
\item Close mapping between $\frac{d\Phi}{dV}$ and $\frac{d\Phi}{dz}$ (FIG. \ref{tds_v_Bz}).

\item Disappearance of the $\frac{d\Phi}{dV}$ and $\frac{d\Phi}{dI}$ signals when the SQUID chip is in contact (FIG. \ref{incontact}, also 1 and 2)

\item Characteristic gate voltage dependence of $\frac{d\Phi}{dV}$ and $\frac{d\Phi}{dz}$ (FIG. \ref{tds_v_Vbg}) and a sign flip of the signals observed in $\frac{d\Phi}{dV}$ (FIG. \ref{tds_v_Vbg} and \ref{imgs_v_Vbg}).

\item Mechanical resonance in the frequency dependence of the $\frac{d\Phi}{dV}$ signal (FIG. \ref{freq_dep})
\end{enumerate}

\fig{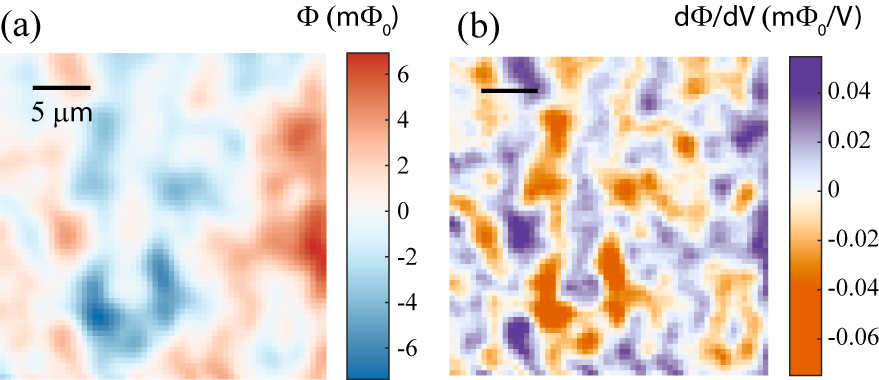}{image_example}
{Scanning SQUID DC flux and $d\Phi/dV$ images of a $EuS/Bi_2Se_3$ bilayer device at $V_{BG}$ = -200 V.}
{(a) A DC flux images ($\Phi$) of as cooled domain structure in a $EuS/Bi_2Se_3$ bilayer. (b)$d\Phi/dV$ image taken simultaneously with the $\Phi$ image. This image is, as we argue in the text, due to electric coupling between the SQUID and sample in the presence of DC magnetic fields. We applied a voltage directly to one terminal of the device with all of the other terminals floating. This scan was taken at $V_{AC}^{pk}$ = 10 V, f = 685 Hz, with the SQUID's chip out of contact with the sample.}{0.7}

All of these effects are easily understandable in the model presented above, and hard to reconcile with a signal that arises due to chiral currents along magnetic domain walls and edges.

Typical DC magnetometry ($\Phi$) and $\frac{d\Phi}{dV}$ images are presented in FIG. \ref{image_example}. We measured the sample as-cooled, with no external field applied at any point during the cooling process. We focused on a 30 $\mu m$ x 30 $\mu m$ area of the device. We found ferromagnetic domain structure in DC magnetometry, the size of which is mostly limited by the SQUID's spatial resolution (FIG. \ref{image_example} a). At large negative  $V_{BG}$ ($-200 V$), we observe a large signal in $d\Phi/dV$. The spatial pattern of $d\Phi/dV$ is very similar to the domain structure observed in the $\Phi$ image (FIG. \ref{image_example} b).

\fig{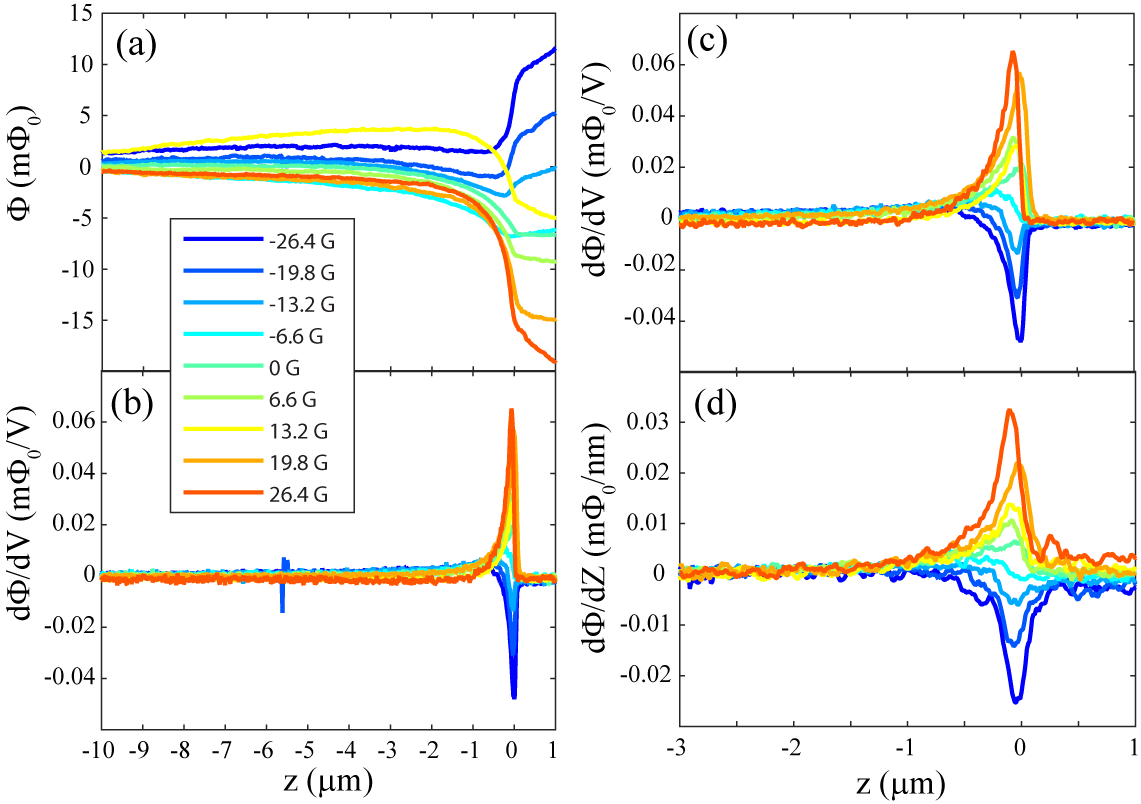}{tds_v_Bz}
{$\Phi$, $d\Phi/dV$, and numerically determined $d\Phi/dz$ height dependencies as a function of applied out of plane magnetic field.}
{(a) $\Phi$ vs. height of the SQUID above the sample. Negative heights indicate the distance above the sample, while positive distances indicate that the SQUID's chip is touching the sample, and does not actually reflect the true height above the sample. $\Phi(z)$ changes significantly as a function of an applied field out of plane field, which occurs due to a combination of the paramagnetic response of the sample and changing domain structure. (b) $d\Phi/dV$ as a function of height. We find that similar to $\Phi(z)$, $d\Phi/dV$ (z) changes strongly with applied field. (c) A zoomed in plot of (b). (d) Numerically determined $d\Phi/dz$ curves obtained from the $\Phi(z)$ curves (see main text). We find that the dependence of $d\Phi/dV(z)$ and $d\Phi/dz(z)$  are qualitatively very similar, as expected from electric coupling. Specifically, the magnetic field at which the touchdown curves change sign,  the relative heights of the curves, and the strong reduction of the signals when the SQUID chip is in contact with the sample all point toward the electric coupling explanation.}{0.9}

To investigate the signal in a different way, we measured the dependence of $\Phi$ and $d\Phi/dV$ on height above the sample (FIG. \ref{tds_v_Bz} a \& b). At a fixed gate voltage and temperature ($V_{BG} = -200 V$ \& $4.2 K$) we measured $\Phi(z)$ and $d\Phi/dV(z)$ and changed the applied out-of-plane field ($B_z$). $\Phi(z)$ (FIG. \ref{tds_v_Bz} a) is tuned by the field by changes in the domain structure and the paramagnetic response of the sample. We found that $d\Phi/dV(z)$ appears to be much more sharply varying than $\Phi(z)$ and that $d\Phi/dV$  goes to zero when the SQUID is in contact with the sample.  

The electric coupling term in Eqn. 5 is directly proportional to $\partial\Phi/\partial z$, so we took the numerical derivative of $\Phi$ with respect to height in order to compare the two. We applied a smoothing filter (over z$\sim 0.1 \mu m$) to $\Phi$ before taking a numerical derivative, in order to more clearly see the qualitative features. We found that the shape and dependence on field of $d\Phi/dz(z)$ qualitatively matches that of $d\Phi/dV(z)$ (Fig. \ref{tds_v_Bz} c \& d ). Specifically, both are significantly sharper than $\Phi$ and are strongly reduced while in contact with the sample.

\fig{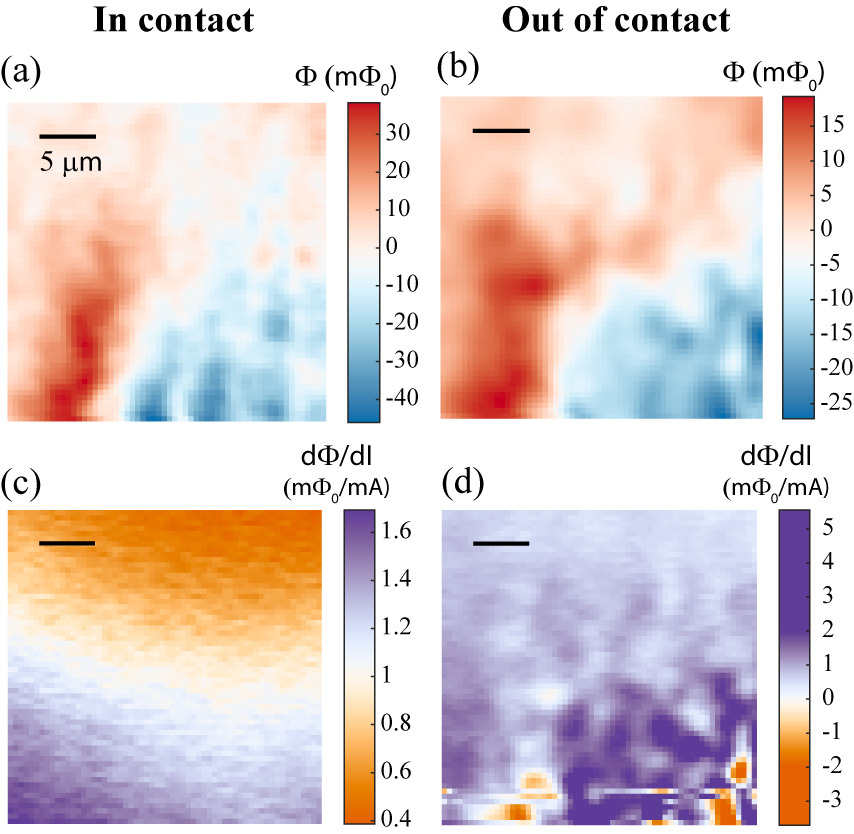}{incontact}
{$\Phi$ and $d\Phi/dI$ images with the SQUID's chip in and out of contact with the sample.}
{(a,b) $\Phi$  images in contact and out of contact, respectively. In contact (a),  the ferromagnetic domain structure was more strongly resolved due to reduced spread of the magnetic fields originating from the sample. (c,d). $d\Phi/dI$ images in and out of contact, respectively. In contact (c), we found that the current flow image was smoothly varying and did not match the domain pattern, consistent with bulk current flowing in the sample. Out of contact (d), the magnetic domain like features appear in $d\Phi/dI$ image, washing out the bulk current signal. The stark difference between $d\Phi/dI$ (and also $d\Phi/dV$) images taken in and out of contact are only easily understandable if the features are due to electric coupling.}{0.7}

The sharp height dependence of $d\Phi/dV(z)$ and its similarity to corresponding $d\Phi/dz(z)$ curves are suggestive that electric coupling is responsible for the observed signal. The disappearance of $d\Phi/dV$  when the SQUID is in contact with the sample is, we believe, a strong piece of evidence that the $d\Phi/dV$ signal is indeed the electric coupling artifact. To further investigate this, we imaged of $d\Phi/dI$ in and out of contact (FIG. \ref{incontact}). 

When the SQUID's chip is in contact with the sample, we expect the electrically-induced vibration of the SQUID to be strongly damped due to the restoring force of the sample itself. We imaged $\Phi$ both in contact and out of contact (FIG. \ref{incontact} a \& b). The only difference between in contact and out of contact $\Phi$ is the sharpness of images, which is because magnetic field lines spread as distance from the sample increases. The $d\Phi/dI$ images both in contact and out of contact (FIG. \ref{incontact} c \& d) are qualitatively different. Out of contact, (FIG. \ref{incontact} d), we primarily observed a similar structure to what is observed in $d\Phi/dV$ (FIG. \ref{image_example} b). However, when we imaged $d\Phi/dI$ in contact (FIG. \ref{incontact} c) the magnetic domain-like features completely disappear. We were left only with a plane-like spatial dependence, consistent with homogeneous current flow in the device. 

The stark difference in the measured $d\Phi/dI$ in and out of contact makes sense if the magnetic-domain like signals in $d\Phi/dI$ are due to electric coupling. In $d\Phi/dV$ the magnetic domain-like feature disappear in contact as well (as we showed in height dependence, FIG. \ref{tds_v_Bz}). Additionally, the apparent lack of any domain-like signal while the SQUID is in contact shows that there is no measurable signal from chiral currents in this particular image.

Another explanation for the in contact behavior could be that the voltage applied to the sample is shorted to the SQUID. We checked this possibility by putting the SQUID in contact with the sample while applying a DC voltage and measuring leakage. We did not observe any leakage current from the sample to ground, indicating that shorting was not an issue in this measurement. The presence of an insulating layer on top of the sample ($AlO_x$) makes shorting unlikely. In previous experiments on top-gated samples, we have observed shorting of a couple of volts to the SQUID, and in these cases we were unable to successfully take SQUID images due to the shorting. These problems were not present in $EuS/Bi_2Se_3$ measurements, indicating that shorting was most likely not occurring while imaging.

\fig{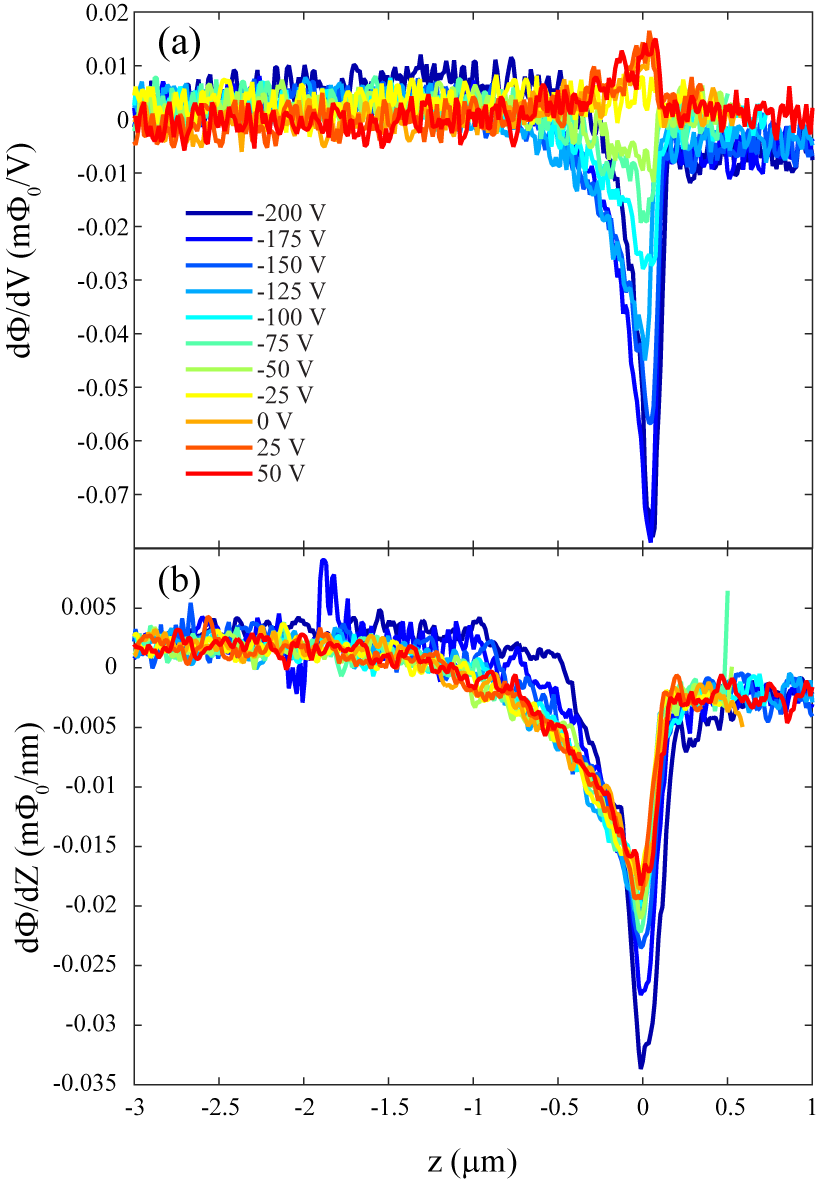}{tds_v_Vbg}
{$d\Phi/dV$ and $d\Phi/dz$ as a function of back gate voltage.}
{This data was taken in an applied field of $B_z$ = -26.4 G, $V_{AC}^{pk}$ = 10 V, $f_{AC}$ = 271.06 Hz.  (a) $d\Phi/dV(z)$ as a function of back gate voltage. We find that $d\Phi/dV(z)$ changes sign at approximately $V_{BG} ~ -25 V$, consistent with a sign flip around zero volts from our basic model of the electric coupling. We have subtracted a constant offset from the data far away from the sample, which became nonzero at positive gate voltages. (b) $d\Phi/dz(z)$ curves as a function of gate voltage. We took the numerical derivative of $\Phi(z)$ and found that the sign does not switch when the sign of $d\Phi/dV(z)$ changes. This means that $dz/dV$ must change sign if the $d\Phi/dV$ signal is from electric coupling, which is predicted by our electric coupling model.}{0.5}

We compared $d\Phi/dV(z)$ and $d\Phi/dz(z)$ as a function of $V_{BG}$ to further corroborate the electric coupling explanation. In order to maximize the observed signal, we applied an out of plane field $B_z$ = -26.4 G. Looking back at Eqn. 5, we expect the electrically-induced vibration, and therefore the electric coupling term in $d\Phi/dV$ to change sign with $V_{BG}$. We measured $d\Phi/dV$ and $d\Phi/dz$ as a function of $V_{BG}$ (FIG. \ref{tds_v_Vbg}) and found that while $d\Phi/dV$ varied strongly and ultimately flipped sign at positive $V_{BG}$, $d\Phi/dz(z)$ varied considerably less. The relative sign between $d\Phi/dz(z)$ and $d\Phi/dV(z)$ changed between $V_{BG}$ = 0 and -25 V., which is consistent with what we predicted in Eqn. 5. The fact the sign of $d\Phi/dV(z)$ changes at a non-zero $V_{BG}$ indicates that the heterostructure's flat band voltage is non-zero, which is plausible.

\fig{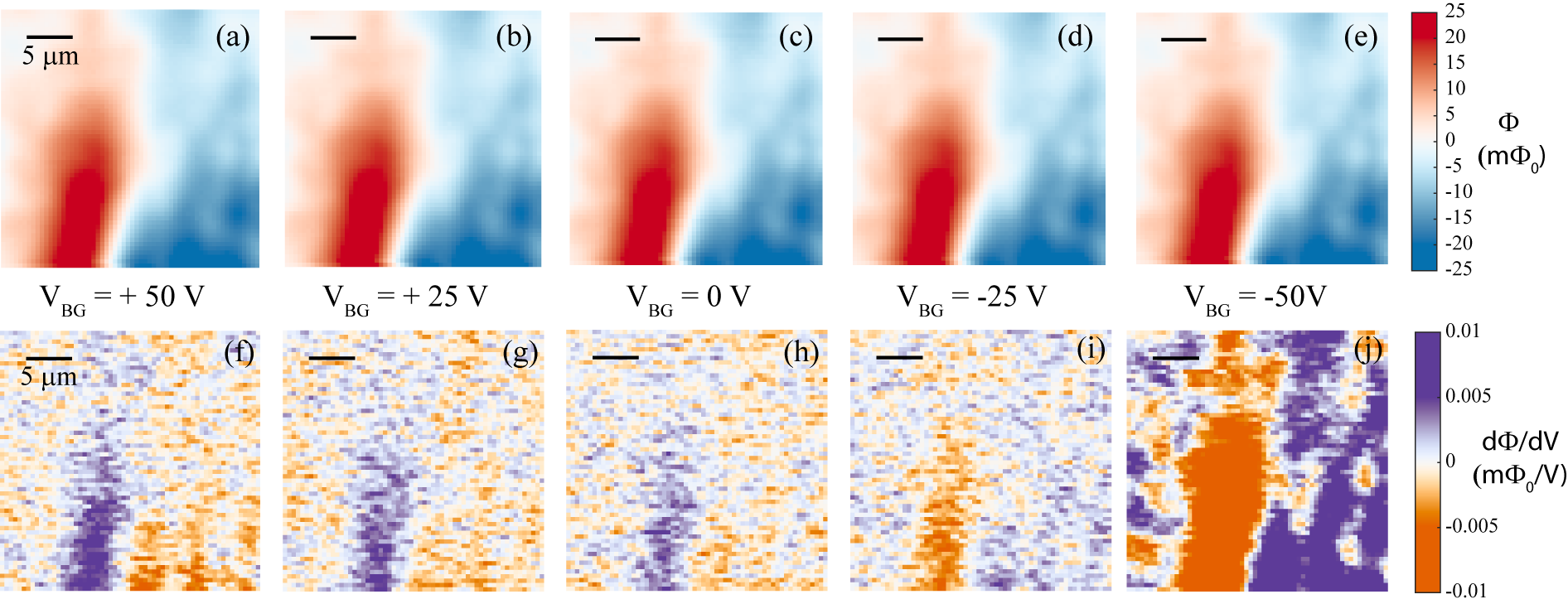}{imgs_v_Vbg}
{$\Phi$ and $d\Phi/dV$ images as a function of gate voltage show a sign flip of the features observed in $d\Phi/dV$ between VBG = 0 and -25 V. }
{Images were taken at f=612.4 Hz, $V_{AC}^{pk}$ = 10 V, $B_z$= 0 G. (a)-(e) $\Phi$ images of ferromagnetic domains as a function of back gate voltage, showing little dependence. (f)-(j) $d\Phi/dV$ images as a function of gate voltage, showing that the relative sign of features in $d\Phi/dV$ change sign as a function of gate voltage. }{0.9}

The sign change of $d\Phi/dV$ is predicted by electric coupling, but inconsistent with a chiral current explanation. We further showed this by taking images at many $V_{BG}$ around zero voltage in zero applied field (FIG. \ref{imgs_v_Vbg}). We found that the DC magnetic features remained constant (FIG. \ref{imgs_v_Vbg} a-e), while the relative sign of the features in $d\Phi/dV$ flipped between $V_{BG}$ = 0 and -25 V. As grown $Bi_2Se_3$ is strongly n-doped, so the presence of any chiral current features near $V_{BG}$ = 0 V is highly unlikely, but the sign flip of the features is inconsistent with a signal from chiral currents.

\fig{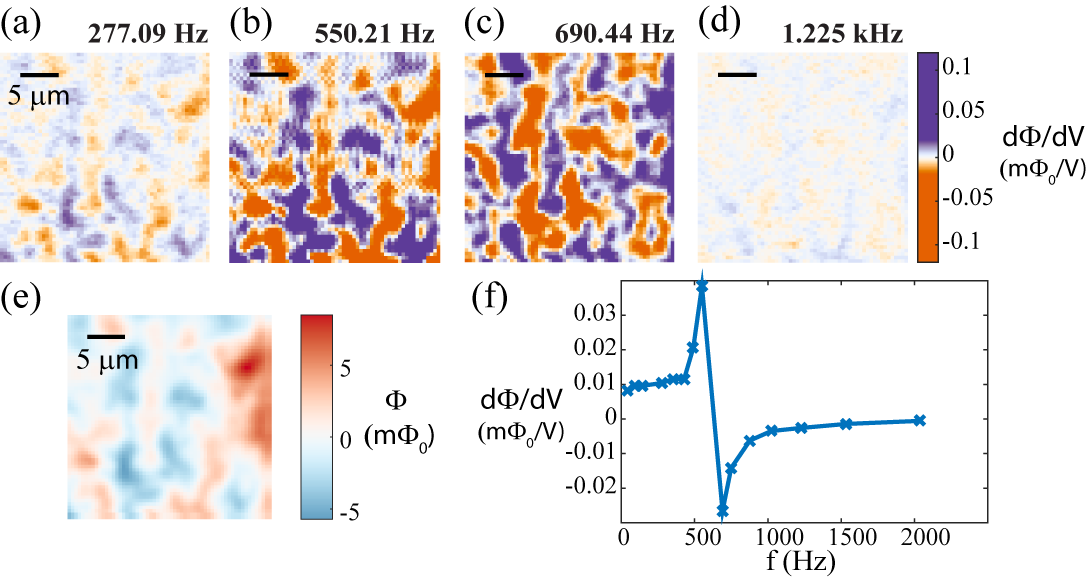}{freq_dep}
{Frequency dependence of $d\Phi/dV$ images shows a mechanical resonance.}
{(a)-(d) $d\Phi/dV$ images as a function frequency show that both the amplitude and sign of the features in the images change with frequency. The sign change happens around 600 Hz, consistent with the calculated resonance expected for the Cu cantilever on which the SQUID sits. (e) $\Phi$ image of the same area for reference, it does not change with frequency of the applied voltage. (f). The measured in phase $d\Phi/dV$ signal at a fixed point on the scans (bottom left corner) as a function of frequency. The sign change and amplitude dependence with frequency closely matches that of a mechanical resonance. }{1.0}

Finally, we also observed that the features in $d\Phi/dV$ flipped sign as a function of the frequency of the voltage excitation (FIG. \ref{freq_dep}). The features in $d\Phi/dV$ images (FIG. \ref{freq_dep} a-d) match the domain structure observed in $\Phi$ (FIG. \ref{freq_dep} e), however the features in $d\Phi/dV$ flipped sign around f=600 Hz. To further show the frequency dependence, we plotted the $d\Phi/dV$ value in the bottom left corner of this area at many frequencies (FIG. \ref{freq_dep} f). We found $d\Phi/dV$ signal's sign flipped and its magnitude peaked at f$\sim 600$ Hz.

The frequency dependence of $d\Phi/dV$ strongly indicates a mechanical resonance of the SQUID's cantilever. In our derivation of the electric coupling we assumed static forces, and therefore did not take any mechanical resonance of the cantilever into account. A resonance leads to a frequency dependence which would also affect the magnitude and sign of $dz/dV$ and therefore the electric coupling term in $d\Phi/dV$. We roughly estimated the resonance of our Cu cantilever using Copper's room temperature values for the Young's modulus (E = 130 GPa) and density ($\rho = 9 g/cm^3$), and approximate size of the cantilever's thickness and length ($T= 100 \mu m$ \& L = 10 mm). We calculated $f_0 \sim$ 600 Hz, close to the measured resonance.

We have performed careful height, field, frequency, and $V_{BG}$ dependencies of $d\Phi/dV$ and $d\Phi/dI$ and compared them to predictions for the derived trivial electric coupling term. The disappearance of the magnetic-domain like signal in $d\Phi/dV$ and $d\Phi/dI$ in contact, and the sign flips of the observed signals as a function of frequency and $V_{BG}$ are all easy to explain by electric coupling and extremely hard to reconcile with a signal from chiral currents along domain walls.

\section{Implications for Ref. \cite{Wang2015}}\label{implications}

We have demonstrated in the previous section that electric coupling is completely responsible for the observed $d\Phi/dV$ and $d\Phi/dI$ images in a new measurement on $EuS/Bi_2Se_3$ bilayers. We will now analyze the images obtained in the studies that led to Ref. \cite{Wang2015} and argue that they are also caused by electric coupling, rather than chiral currents. 

\fig{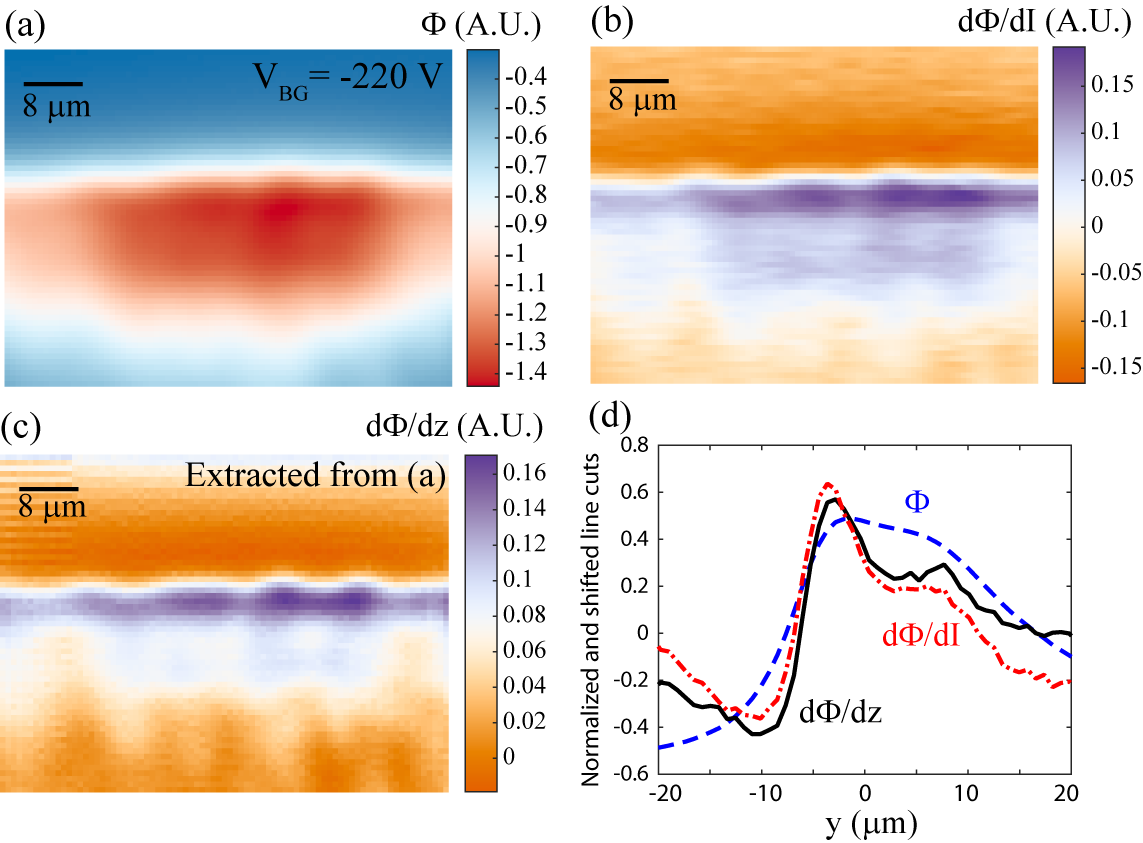}{edge_dPhidz}
{$d\Phi/dI$ edge image from Ref \cite{Wang2015} closely match extracted $d\Phi/dz$ images.}
{(a) (same as FIG. 2B from \cite{Wang2015}). Magnetic flux image ($\Phi$) of the edge of a sample which has been cooled in an out of plane field. (b) (same as FIG. 2E from \cite{Wang2015}) Measured magnetic response from current ($d\Phi/dI$) taken simultaneously, showing what qualitatively matches the expected signal from edge currents. (c) Extracted $d\Phi/dz$ from height propagation of (a), see the main text for details. $d\Phi/dz$  reproduces the features observed in (b). (d) Normalized and shifted vertical line cuts of (a-c), showing the shape of $d\Phi/dI$ (red dot dashed line) more closely matches $d\Phi/dz$ (black solid line) than $\Phi$ (blue dashed line).}{0.75}

We will first show that the $d\Phi/dI$ images from Ref. \cite{Wang2015} match very closely the spatial dependence expected for the electric coupling artifact ($d\Phi/dz$) by propagating the $\Phi$ images in height. Specifically, we have reproduced Fig. 2 B \& E from Ref. \cite{Wang2015} in FIG. \ref{edge_dPhidz} a \& b. 

In order to extract $d\Phi/dz$ from a single magnetometry image, we utilized Fourier techniques to propagate the magnetic image up in height:

\begin{equation}
\Phi (z=h+dh) = IFFT [ e^{-k dh} FFT [\Phi(z=h)] ]
\end{equation}

Where FFT and IFFT are the 2D fast Fourier transform and its inverse, k is the magnitude of the spatial frequency $\sqrt{k_x^2+k_y^2}$, $h$ is the height of the SQUID above the sample, and $dh$ is the height we have propagated the magnetic image \cite{roth}.

In order to get $d\Phi/dz(h)$, we then subtract the two images:

\begin{equation}
\frac{d\Phi}{dz}(h) \approx \frac{\Phi(h) - \Phi(h+dh)}{dh}
\end{equation}

We found that the extracted $d\Phi/dz$ image (FIG. \ref{edge_dPhidz} c) is nearly identical to the spatial structure in the measured $d\Phi/dI$ image (FIG. \ref{edge_dPhidz} b). This is strongly suggestive evidence that the observed signal in Ref. \cite{Wang2015} is due to electric coupling.

From simulations (see FIG. \ref{edge_sim} \& \ref{domain_sim}), we found that the spatial dependence of a  chiral current signal more closely matches the spatial dependence of the magnetic fields from an out of plane domain, rather than its height derivative. Specifically, for out-of-plane domains, the magnetic fields from a fully polarized domain are equivalent to a current flowing along the edge of the domain. This is a generalization of $\mu = I a$, where $\mu$ is the magnetic dipole moment of a small dipole, and $I$ is the current flowing in a loop of area $a$. 

The spatial dependence of the $d\Phi/dI$ image presented in Ref. \cite{Wang2015} is nearly identical to what we expect for electric coupling ($d\Phi/dz$), and is sharper than the magnetic fields we naively expect for chiral currents along the edge of an out of plane polarized domain structure. Line cuts of the images (FIG. \ref{edge_dPhidz} d) further show the similarity between the simulated $d\Phi/dz$ and $d\Phi/dI$.  

Next, we will look at unpublished data that was taken in the same cooldowns as the results presented in Ref. \cite{Wang2015}. We show that two of the main pieces of evidence for electric coupling in the new measurements are also present in those data sets. Specifically, the $d\Phi/dI$ images observed in Ref. \cite{Wang2015} also disappear while the SQUID is in contact, and signals of reversed sign are observed at large positive back gate voltages for written domains.

\fig{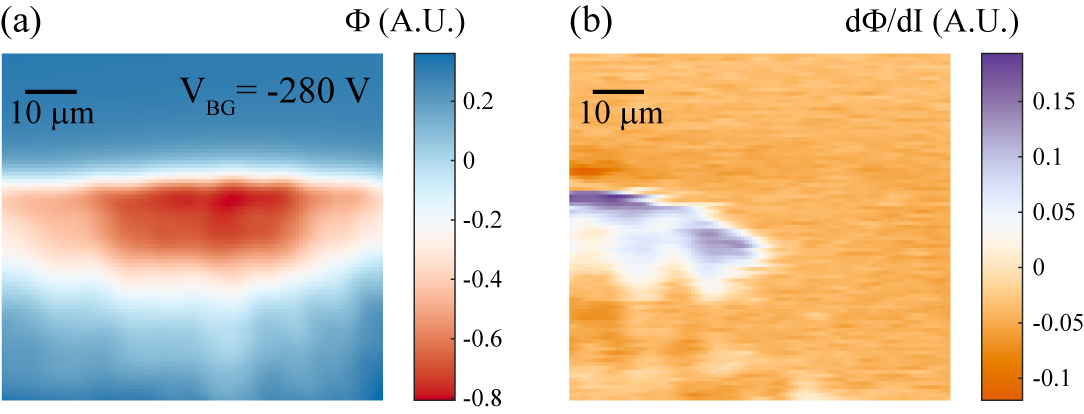}{incontact_edge}
{$d\Phi/dI$ edge image taken in the same gate series presented in Ref \cite{Wang2015} shows the signal disappearing in contact.}
{(a) Magnetic flux image ($\Phi$) of the edge of a sample which has been cooled in an out of plane field. Both images were taken at $V_{BG}$= -280 V. (b) Measured magnetic response from current ($d\Phi/dI$) taken simultaneously, showing what looks like edge currents disappearing in the right half of the image. This is most likely because the SQUID is touching the sample for the right half of the image due to a scan plane error, and the signal disappears in contact. The lack of any signal in the right half of the image puts  strong limits on the real chiral current signal present at this gate voltage. It is less than 5\% of the signal observed when the SQUID was out of contact.} {0.75}

Taken in sequence with the other gate voltage images shown in Fig. 2 of \cite{Wang2015}, an image was taken at $V_{BG}$ = -280 V, where $d\Phi/dI$ goes to zero for a large portion of the scan (FIG. \ref{incontact_edge}). This disappearance of $d\Phi/dI$ during the scan was previously interpreted as an inhomogeneous back gate voltage distribution in the sample before an equilibrium of charge distribution was reached. However, a more likely explanation for this effect is that the signal disappears in contact, and that the SQUID is touching the sample for the right half of the image.

The right side of FIG. \ref{incontact_edge} b shows that while the SQUID is contact, there is no observable signal indicating an edge current. This indicates that not only is the electric coupling artifact present in these measurements, but that it accounts for all of the signal observed in the left half of the scan.

In combination with the effects described in previous section, we have also observed in multiple cooldowns that the voltage applied to the piezo-positioner which is required to touch the sample with the SQUID varies as a function of $V_{BG}$. We found that the SQUID seems to 'snap to' the sample at large negative gate voltages. In other words, a DC electric force leads to height variation for images taken with nominally the same parameters.  We believe that this effect, in combination with the disappearance of the signal when the SQUID is in contact, can explain the lack of a signal observed at $V_{BG} = -350 V$. Images of the same sample, in the same cooldown, show strong signals in $d\Phi/dI$ at a similar back gate voltage ( $V_{BG} = -357 V$), showing that it is likely that Fig.2F from Ref. \cite{Wang2015} was taken in contact with the sample. 

\fig{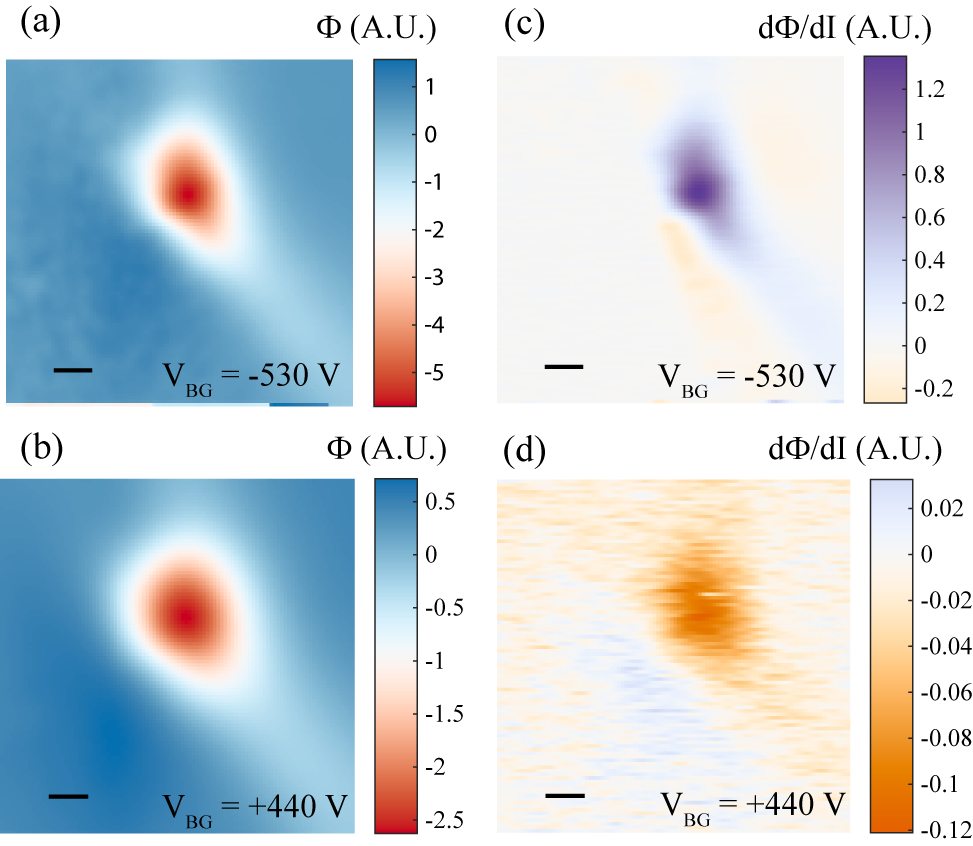}{plus_minus}
{$d\Phi/dI$ images taken over domain structure written by the field coil (similar to Fig. 4 from Ref. \cite{Wang2015} ) show a sign reversal between large positive and negative gate voltages}
{(a,b) Magnetic flux images ($\Phi$) of field coil written domain structure at $V_{BG}$ = -530 V and +440V, respectively. The sample appears to be farther away in (b). (c,d) Measured magnetic response from current ($d\Phi/dI$) taken simultaneously with (a,b), respectively.  The sign of the main feature changes from positive to negative as the sign of the gate voltage is changed. This is consistent with the electric coupling explanation, specifically the prediction of a sign flip of the electric coupling artifact in Eqn. 5. Here we have presented images taken with the voltage applied to different ends of the sample. The magnitude of the signals with the applied voltage to the same side of the sample are approximately a factor of 20 weaker at positive gate voltages than at negative gate voltages. We attribute the difference in the size of the $d\Phi/dI$ mainly to the effects of screening at positive gate voltages and $V_{BG}$ dependent scan height (see text). } {0.75}

A sign reversal of the $d\Phi/dI$ signal was observed between large negative and positive gate voltages in unpublished measurements by Wang, et al. on domain structure written with the field coil (FIG. \ref{plus_minus}). Domain structure was written by applying an inhomogeneous DC magnetic field with the field coil, similar to what was done for Fig. 4 in Ref. \cite{Wang2015}. The sign of the $d\Phi/dI$ signal switched between large negative and positive back gate voltages (FIG. \ref{plus_minus} c \& d). This is inconsistent with a signal from chiral currents, and can be explained by the electric coupling artifact.

The $d\Phi/dI$ signal at large positive $V_{BG}$ was significantly weaker than at large negative $V_{BG}$, roughly a factor of 20 for bias voltages applied to the same end of the sample. The asymmetry in signal size is not predicted by our simple model. Multiple factors can in principle contribute to the asymmetry, including height differences between the scans, differences in the actual applied local voltage due to contact resistance, differences in the screening length, and the contribution of a true chiral current signal at negative gate voltages. The asymmetry is also observed at lower gate voltages in the new measurements, where we have ruled out the presence observable chiral current signals by scanning in contact. Therefore we believe the asymmetry does not require the presence of chiral currents. The large difference in signal size can be accounted for by taking into account screening and height.

For Fig. \ref{plus_minus}, we estimate that the difference in size of $d\Phi/dz$ due to $V_{BG}$ induced height differences between negative and positive gate voltages is $\sim$3 by the looking at calculated $d\Phi/dz$ images.  In Fig. \ref{imgs_v_Vbg}, we observed a factor of 5 difference between positive and negative gate voltages. At low voltages, we do not observe a large height shift in $\Phi$ images (FIG. \ref{imgs_v_Vbg} a-e), so we attribute this factor of 5 mainly to screening effects. Therefore, height and screening effects contribute a lower bound of a factor of $\sim$15 to the asymmetry between positive and negative gate voltage $d\Phi/dI$ images, which is close to the observed factor of $\sim$20.

We have shown that the signal observed in Ref. \cite{Wang2015} is consistent with a signal that is due to electric coupling between the SQUID and the $EuS/Bi_2Se_3$ device. The measured current image $d\Phi/dI$ at the edge of the sample very closely matches the extracted $d\Phi/dz$ image, and is naively 'too sharp' to match the expected spatial variation expected for chiral currents. We also found evidence that the signal in unpublished measurements done in the same experiments as Ref. \cite{Wang2015} disappeared when the SQUID was in contact, and flipped sign between negative and positive back gate voltages.

Taken together with the electric coupling model developed in this paper and the new measurements performed on $EuS/Bi_2Se_3$, we have shown that the signals reported in Ref. \cite{Wang2015} are dominated by electric coupling and show no clear evidence of chiral currents.

\FloatBarrier
\section{Electric coupling in other current imaging experiments} \label{other_works}
\FloatBarrier
Scanning SQUID has been utilized to measure $d\Phi/dI$ and reconstruct current densities in $LaAlO_3/SrTiO_3$, HgTe quantum wells, and InAs/GaSb quantum wells (\cite{kalisky,nowack,spanton}). In HgTe and InAs/GaSb, we observed current along the edges of the devices for certain top gate voltages. In $LaAlO_3/SrTiO_3$, we observed that the more current flowed along features due to the tetragonal domain structure of STO. The electric coupling effect described above cannot explain the main observations of those papers for a number of reasons. 

Most important, there were no features in $\Phi$ which match the observed current features in any of the previous scanning SQUID results. The electric coupling artifact produces features in $d\Phi/dV$ or $d\Phi/dI$ which qualitatively match the signal observed in $\Phi$. If the samples are non-magnetic ($\Phi$ = 0, and therefore $d\Phi/dz$ = 0), then the electric coupling term in Eqn. 5 also goes to zero. In both HgTe and InAs/GaSb, there were no observed features in $\Phi$. In $LaAlO_3/SrTiO_3$, some of the observed samples were magnetic, as discussed in \cite{Bert}, however this magnetism occurred in resolution-limited magnetic dipoles, which does not match the quasi 1D features observed in $d\Phi/dI$.

Additionally, in both HgTe and InAs/GaSb the gates are top gates rather than back gates, which significantly changes the electric field environment. Both top and back gates modulate the charge on the device itself, but the metallic top gate is between the grounded SQUID and the sample, which will strongly screen the charge on the sample from the SQUID, resulting in an extreme reduction electric coupling. 

\section{Conclusion}

In conclusion, electric coupling between the SQUID and sample can lead to an electric coupling artifact in a specific type of measurement and sample. Specifically, $d\Phi/dI$ and $d\Phi/dV$ measurements of back-gated semiconducting samples with magnetic features. Any SQUID measurements of this type should be very strongly scrutinized. The electric coupling effect is present and the dominant signal in our new measurements of $EuS/Bi_2Se_3$, and can also explain all of the observations of Ref. \cite{Wang2015}.

If the sample is non-magnetic or top-gated, or the measurements are not $d\Phi/dI$ or $d\Phi/dV$, the electric coupling effect described here will not be relevant.

\section{Acknowledgments}
We acknowledge Christopher Watson for useful discussions. These measurements and analysis were supported by the Department of Energy, Office of Science, Basic Energy Sciences, Materials Sciences and Engineering Division, under Contract DE-AC02-76SF00515. Earlier measurements were supported both by FAME, one of six centers of STARnet, a Semiconductor Research Corporation program sponsored by MARCO and DARPA, and by the Department of Energy, Office of Science, Basic Energy Sciences, Materials Sciences and Engineering Division, under Contract DE-AC02-76SF00515. The SQUID microscope and sensors were developed with support from NSF-NSEC 0830228 and NSF IMR-MIP 0957616. Y.H.W. was partially supported by the Urbanek Fellowship of the Department of Applied Physics at Stanford University. For sample preparation, F.K., P. J-H., and J.S.M. would like to thank support by the MIT MRSEC through the MRSEC Program of the National Science Foundation under award number DMR-0819762. Partial support for sample development was provided by NSF (DMR-1207469), ONR (N00014-13-1-0301) (F.K. and J.S.M.) and by the DOE, Basic Energy Sciences Office, Division of Materials Sciences and Engineering, under Award No. DE-SC0006418 (F.K. and P. J-H.).

\FloatBarrier

\bibliographystyle{ieeetr}
\bibliography{bibliography.bib}
\end{document}